\documentclass[
      preprint,
      superscriptaddress]{revtex4-1}

\usepackage{bbold}
\usepackage{xcolor}
\usepackage{braket}
\usepackage{subfig}
\newcommand{\sinc}{{\rm sinc}}

\newcommand{\sign}{{\rm sign}}

\usepackage{amsmath,amssymb}
\usepackage{graphicx}% Include figure files
\usepackage{dcolumn}% Align table columns on decimal point
\usepackage{bm}% bold math
\usepackage[latin1]{inputenc}
\usepackage{float}
\usepackage{color}
\usepackage{booktabs}
\definecolor{rojo}{rgb}{1,0,0}
\definecolor{verde}{rgb}{0,0.8,0.5}
\definecolor{azul}{rgb}{0,0,1}
\definecolor{rosa}{cmyk}{0,1,0,0}
\usepackage[toc,page]{appendix} 
\usepackage{latexsym}
\usepackage{mathrsfs}
\usepackage{bm}
\usepackage{amsthm}
\usepackage{adjustbox}
\usepackage{bbm}
\usepackage{amsfonts}
\usepackage{amssymb}
\usepackage{epsfig}
\usepackage{hyperref}
\hypersetup{colorlinks=true, citecolor=blue}
\usepackage{soul}
\usepackage[normalem]{ulem}
\usepackage[cmtip,all]{xy} 
\usepackage{comment}
\usepackage{multirow}
\usepackage{mathtools}
\newcolumntype{L}{>{$}l<{$}}
\newcommand{\longsquiggly}{\xymatrix{{}\ar@{~>}[r]&{}}} 

\begin{document}

\title{Emergence of Floquet edge states in the coupled Su-Schrieffer-Heeger model}

\author{Carla Borja}
\affiliation{School of Physical Sciences and Nanotechnology, Yachay Tech University, Urcuqu\'i 100119, Ecuador}
\author{Esther Guti\'{e}rrez}
 \affiliation{Escuela Superior Polit\'ecnica del Litoral, ESPOL, Departamento de F\'isica, Facultad de Ciencias Naturales y Matem\'aticas, Campus Gustavo Galindo
 Km. 30.5 V\'ia Perimetral, P. O. Box 09-01-5863, Guayaquil, Ecuador}	
\author{Alexander L\'{o}pez}
\email[To whom correspondence should be addressed. Electronic
address: ]{alexlop@espol.edu.ec}
\affiliation{Escuela Superior Polit\'ecnica del Litoral, ESPOL, Departamento de F\'isica, Facultad de Ciencias Naturales y Matem\'aticas, Campus Gustavo Galindo
 Km. 30.5 V\'ia Perimetral, P. O. Box 09-01-5863, Guayaquil, Ecuador}	

%\date{\today}

\begin{abstract}
The emergence of non equilibrium topological phases in low dimensional systems offers an interesting route for material properties engineering. We analyze the dynamical modulation of two coupled one-dimensional chains, described by the Su-Schrieffer-Heeger model. We find that the interplay of driving interactions and interchain coupling leads to the emergence of non-equilibrium edge states with nontrivial topological properties. Using an effective Hamiltonian approach, we quantify the emergent topological phases via the winding number and show that oscillations in the mean pseudospin polarization arise as a consequence of the periodic modulation. The patterns of these pseudospin oscillations are different for the static trivial and topological phases offering a dynamical means to distinguish both physical configurations. The system also exhibits non integer values of the winding number, which have been recently reported experimentally in connection to spin textures.
\end{abstract}
\maketitle

%----------------------------------------------------------------%
\section{Introduction}\label{section1}
Since the pioneering work of Von Klitzing\cite{VonKlitzing1980} on the quantum Hall effect (QHE), there have been a wealth of proposals for the observation of physical states of matter with so called topological properties\cite{Haldane1988,Kane2005,Bernevig2006,Koenig2007,Hasan2010}. Essentially, these topological materials can support a bulk bandgap but have protected gapless edge states at their boundaries. Typically, the protection is due to either time reversal symmetry and/or spatial inversion symmetries although it has been shown that spinless systems might possess protected edge states in absence of time-reversal symmetry\cite{Bernevig2014}.
Within their proposal, the authors show that the edge states are protected via the symmetries of a given subset of the crystallographic point groups. Recently, many theoretical works have addressed the role of different interactions in the generation of these boundary protected edge states. From the experimental point of view, the first observation of such topological properties in $HgTe/(Hg,Cd)Te$ quantum wells\cite{Koenig2007} paved the road for a plethora of other experimental and theoretical works showing the potential of these materials for technological implementations of nanodevices. 
The experimental realization and the observation of topological phases in two \cite{Koenig2007} 
and three\cite{3DTI} dimensional materials has revealed the technological challenges in observing dissipationless features in the transport regime of these systems. This in turn has motivated a quest for lower dimensional systems with emergent topological nontrivial phases. An experimentally realized one-dimensional system, where nontrivial edge states can occur is the Su-Schrieffer-Heeger (SSH) model, originally proposed to discuss the solitons in the poly-acetylene molecule\cite{ssh-original}. Within this model, non-interacting spinless fermions hop among lattice sites, subject to staggered potentials associated originally to single and double bonding of the poly-acetylene molecule. More interestingly, in the energy spectrum, these edge statesof the finite size sample are related to the emergence of flatbands in the bulk spectrum, meaning that they are localized at the boundaries of the sample. 

\noindent Some realizations of this one dimensional chain have been put forward in references\cite{Platero2013,Nature2013_Kitawaga,Meier2016,acoustic-ssh,Nature2019Xie}.  Further studies have considered the role of interactions in the SSH model. For instance, upon the application of an external static electric field to the one-dimensional SSH model, it is shown in reference \cite{Zak2020} that an anomalous phase dielectric response can be found. In addition, recent works\cite{dal2015floquet, StJean} explore the role of periodically driving interactions in a single chain system and found via the associated non-equilibrium Zak phase\cite{Zak1989} that nontrivial topological phases arise which enrich the associated phase diagram among trivial and nontrivial sectors. This Zak phase is directly related to the winding number which is a well established topological invariant, and intuitively can be understood to characterize the number of times a closed curve encircles a given point in  momentum space. In physical systems, the nontrivial nature of the winding number is directly related to the emergence of states at the boundaries of the sample. Therefore, these emergent states are also termed edge states. 

\noindent Thus, analyzing the emergence of boundary edge states, induced by periodically driving protocols\cite{Grifoni1998}, offers a means to study nontrivial responses of low dimensional systems in the nonequilibrium realm.  
Interestingly, the role of periodic driving has been extensively considered in different physical systems.  In pioneering works,\cite{Oka09,Dora2009,NP2011Lindner,PRB2014PerezPiskunow2,Grushin2014,PRB2017Refael,PRB2018Lindner2,PRL2018Refael,Lu2019} it is shown that nontrivial topology can be induced in static systems by driving fields creating the so-called Floquet Topological Insulators (FTI) which show boundary edge states not achievable within the static scenario. Generally speaking, the topological response of a physical system can be assessed via several topological invariants, such as the Chern number, spin-Chern number $\mathcal{Z}_2$, and higher order extensions. There is indeed a classification for these topological systems which allows one to determine to which class belongs a given system\cite{topoclass0,topoclass1,topoclass2,topoclass3,topoclass4}. 
In driven one-dimensional systems, the Zak phase\cite{dal2015floquet} has been proposed as a well defined topological invariant that serves for characterization of the phase diagram relating trivial and nontrivial phases in the relevant parameter regimes of the model considered. 
These studies rely on the generation of the boundary edge states which encode the nontrivial topological properties of the system under consideration. This Zak phase can be related to the winding number and we focus on the last one for discussing our results.

\noindent Recently, some works \cite{Zhang2017,li2017topological} have theoretically explored the static coupling of two SSH chains and show that the associated phase diagram derived from the winding number calculation, leads to further interesting physical regimes. Actually, the interplay of different interactions is expected to lead to the generation of trivial-nontrivial topological phase transitions as, for instance, it was originally described in the Kane and Mele\cite{Kane2005} proposal for a two-dimensional topological insulator which was based on the interplay of two competing spin-orbit interactions in graphene. Their work was in turn an extension of the parity anomaly proposal previously put forward by Haldane\cite{Haldane1988} who showed that a quantum Hall effect (QHE) could be achieved without the Landau levels required in its original experimental realization\cite{VonKlitzing1980}.  

\noindent The role of electromagnetic radiation in a two chain coupled system, described via the SSH model has been put forward in reference\cite{coupledssh2020}, where the authors show that in addition to the static topologically protected flat bands arising within the gap between conduction and valence, a high frequency radiation field induces finite-energy curved bands inside the gap of subbands.  

\noindent In this work, we study two coupled one-dimensional SSH chains, which are subject to periodic driving, focusing on the interplay among periodic driving and interchain coupling to generate nontrivial topological configurations, assessed via the winding number. The motivation for using a driven gate voltage instead of the radiation field discussed in reference\cite{coupledssh2020} is to offer experimentalists an alternative route to dynamically manipulate the low dimensional topological properties of the system. Thus, by exploring a finite-size system, we obtain the quasi energy spectrum and by analyzing the zero energy edge states we show the emergence of quasi flat bands in the quasienergy spectrum, which are absent in the static counterpart configuration. In order to quantify the emergent topological phases, we derive an effective bulk Hamiltonian within the high frequency regime and give the corresponding quasi energy spectra in momentum space. Then, it is given a characterization of the effective high frequency dynamical topological effects by means of the winding number. It is thus shown that the associated phase diagram allows for additional trivial-nontrivial transitions, as compared to the static scenario. 
This in turn, might lead to feasible proposals for on demand control of the topological phases via periodic driving protocols on one dimensional systems described with the SSH model. Moreover, we find that along one of the chains, the pseudospin polarization, which is related to the occupation differences in the bipartite system,  shows driven oscillations. These driven oscillations can be used to weight the bulk quasi energy spectrum for suitable initially prepared states, providing an indirect quantification of interplay among bandgap opening and difference in sublattice occupation due to the periodic driving protocol. It allows us to distinguish among trivial and nontrivial topological configurations.     

\noindent The structure of the paper is as follows. In section \ref{section2} we present the model and discuss the numerical results for the quasi energy spectra showing the emergence of additional zero energy edge states for a finite size sample. We also present the bulk properties, within the intermediate and high frequency regimes for the quasi energy spectrum, winding number and pseudospin polarization. Then, we discuss our findings and give concluding remarks in section \ref{section3}, arguing how a feasible experimental detection scheme of these findings can be realized within currently available techniques. Finally, the explicit calculations leading to the main results  are given in the appendix \ref{appendix}.
\section{model}\label{section2}
We consider a periodically driven coupled system of two one-dimensional chains with hopping non-interacting spinless electrons. As shown in panel (a) of \figurename{1}, which following \cite{li2017topological}, the static Hamiltonian for the bipartite system is given as
\begin{equation}
H_0=\sum_{j=1}^N wa^\dagger_{j+1}b_j+va^\dagger_jb_{j+1}+ca^\dagger_jb_j+H.c.    
\end{equation}
where $a^\dagger_j$ and $b^\dagger_j$ are the creation operators at sublattice A and B of the $jth$ unite cell, whereas $v$ and $w$ denote the intra and inter cell coupling strengths, respectively and $c$ measures the interchain coupling parameter. For $c=0$, onerecovers the original SSH model, that is extensively studied in the literature as a one dimensional system supporting localized edge states at its boundaries. Interestingly, the SSH Hamiltonian was originally proposed as a theoretical model to describe solitary states in the poly-acetylene molecule \cite{ssh-original}. Upon coupling two of these SSH chains, it has been shown in reference \cite{li2017topological} that, a richer topological structure emerges where the system can be driven between trivial and nontrivial topological phases meaning essentially that, those previously mentioned edge states are present or absent at the system boundaries.
\begin{figure}
    \centering
    \includegraphics[height=6.5cm]{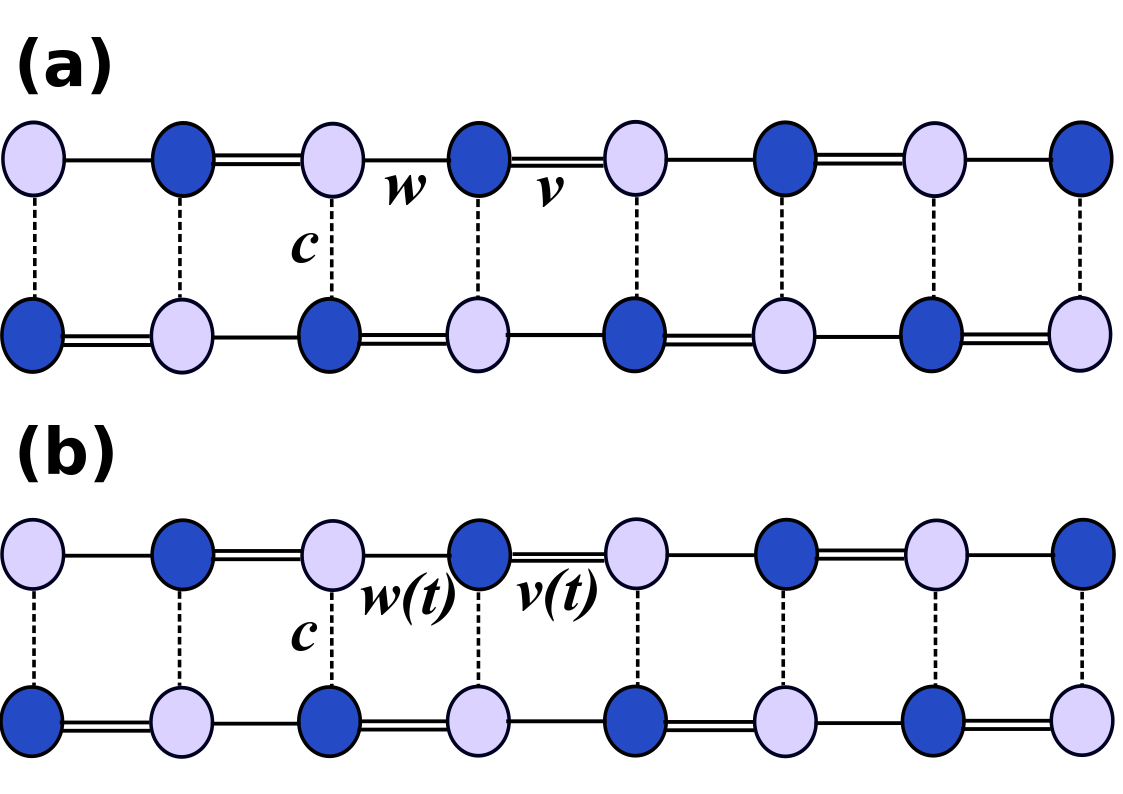}
    \caption{Schematic representation of the two coupled one dimensional SSH chains with hopping non-interacting spinless electrons with interchain coupling strength $c$.  (a) The shaded box represents the unit cell with static intracell (intercell) parameter $v$ ($w$). (b) The system subject to periodic driving with effective time-dependent parameters $v(t)$ and $w(t)$.}
    \label{fig:my_label2}
\end{figure}

\noindent When the two-chain system is subject to a periodic driving protocol, with effective time-dependent parameters $v(t)=v+G(t)$ and $w(t)=w+G(t)$, as shown in panel (b) of \figurename{1}, one can effectively describe the system with time-dependent coupling strengths associated to the periodic contribution $G(t)=G(t+T)$, with $t$ and $T$ being the time variable and period of the perturbation, respectively. For an actual realization of the driving interaction term, we consider the protocol $G(t)=2g\cos\Omega t$, proposed in reference \cite{dal2015floquet} to study non-equilibrium properties of the SSH system, i.e.
\begin{equation}
V(t)=2g\cos\Omega t\sum_{j=1}^{N}( a^\dagger_{j+1}b_j+a^\dagger_jb_{j+1}+h.c.)    
\end{equation}
where $\Omega=2\pi/T$ and $g$ are the frequency and effective driving strength parameters. Thus, one gets an effective time-dependent Hamiltonian periodic in time. We now solve the quasi energy spectrum via an exact numerical diagonalization of the driven problem. We first solve the problem for a finite sized chain by numerically finding the quasi energy spectrum of the system and discuss the resulting photoinduced band spectrum.

\noindent Although the Floquet theory has appeared more often in the literature, mostly thanks to the theoretical proposals and experimental realizations of FTI, we briefly give an account of the main ingredients necessary for dealing with the quasi energy spectrum of the driven system. Hence, we consider a generic time-dependent periodic Hamiltonian $H(t+T)=H(t)$, where $T$ is the time period of the Hamiltonian. From the time-dependent Schr\"odinger equation, this Hamiltonian generates the time evolution $i\partial_t\Psi(t)=H(t)\Psi(t)$.  The solution can be writing as $\Psi(t)=e^{i\epsilon t}\Phi(t)$, with $\Phi(t+T)=\Phi(t)$ being the periodic part of the solution, which is termed the Floquet state and $\epsilon$ is the quasi energy\cite{Grifoni1998}. This quasi energy spectrum stems from the fact that the Hamiltonian is time-dependent, implying that the energy is not a conserved quantity and it is similar to the quasi momentum in spatially periodic systems. In addition, given the periodic nature of the dynamical generator, the quasi energies are also periodic since they are defined modulo $\hbar\Omega$. 

\noindent Upon introduction of the ansatz $\Psi(t)=e^{i\epsilon t}\Phi(t)$ into the time-dependent Schr\"odinger equation, one gets a time independent evolution equation of the Floquet states: $\mathcal{H}\Phi(t)=\epsilon\Phi(t)$, with the Floquet Hamiltonian $\mathcal{H}=i\partial_t-H(t)$. Then, one performs a Fourier expansion and obtains an infinite dimensional eigenvalue equation whose solutions give the aforementioned quasi energy spectrum of the problem. Interestingly, the main physical features of a periodically driven system are obtained from the Floquet spectrum and Floquet states. 

\noindent Thus, we now discuss the solutions of the quasi energy spectrum for the driven coupled SSH model, first discussing the finite size sample with $N=10$ sites, corresponding to $20$ dimers, which is a long enough sample to get the main results reported here. Following reference\cite{dal2015floquet}, we have restricted the solution to the first harmonic corrections to the static Hamiltonian. These harmonics correspond to the Fourier expansion of the periodic Hamiltonian in the frequency realm and are also term side bands since, as it was described in the description of the Floquet theory, they are connected to the periodic nature of the quasi energy spectrum. We stress that restriction to first harmonic contributions gives the correct physical picture in the high frequency regime since the the modes get uncoupled for this off resonant scenario. As expected, at intermediate and low frequencies higher order harmonics should be included as their contribution becomes more relevant to modify the physical properties of the system.
In order to differentiate the coupled chains from the uncoupled situation, we have set the interchain coupling term $c=w$ to model strongly coupled chains. In addition, the amplitude of the driving is set to $g=0.25c$ and the frequency value to $\hbar\Omega=4c/3$.
\noindent  Hence, we consider \figurename{2}. Panel (a) and (b) show the quasi energy spectrum without edge states of the undriven Hamiltonian described in reference\cite{li2017topological}, for the parameter configurations $v=w$ and $w=v+c$ respectively. For the undriven coupled chains, Li et al\cite{li2017topological} derived the effective static bulk Hamiltonian (see next section), and establish the conditions for topologically trivial and nontrivial static phases. For this purpose, they define 

\begin{eqnarray}
x&=&(v+w)\cos k+c\\\nonumber
y&=&(v-w)\sin k
\end{eqnarray}
from which the winding number is obtained as
\begin{equation}
\mathcal{N}=\frac{1}{2\pi}\int_c\frac{xdy-ydx}{r^2}     
\end{equation}
Upon integration, the winding number reads 
\begin{equation}
\mathcal{N}= 
\begin{cases}
0,&|v+w|<c\\
    \sign{(v^2 -w^2)},& \text{otherwise}
\end{cases}
\end{equation}
\begin{figure}[ht]
    \centering
\subfloat[]{\includegraphics[height=6.5cm]{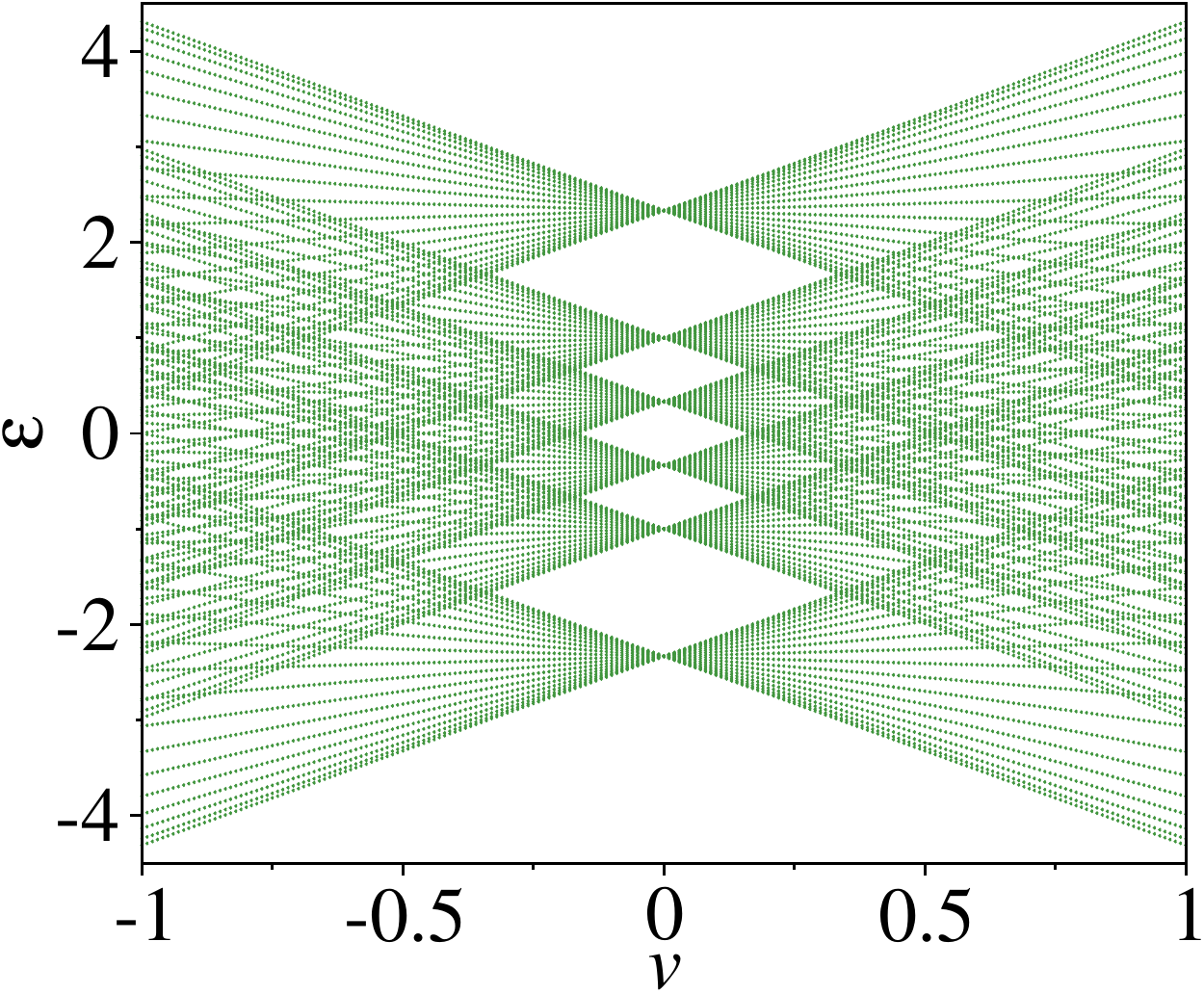}}
\subfloat[]{\includegraphics[height=6.5cm]{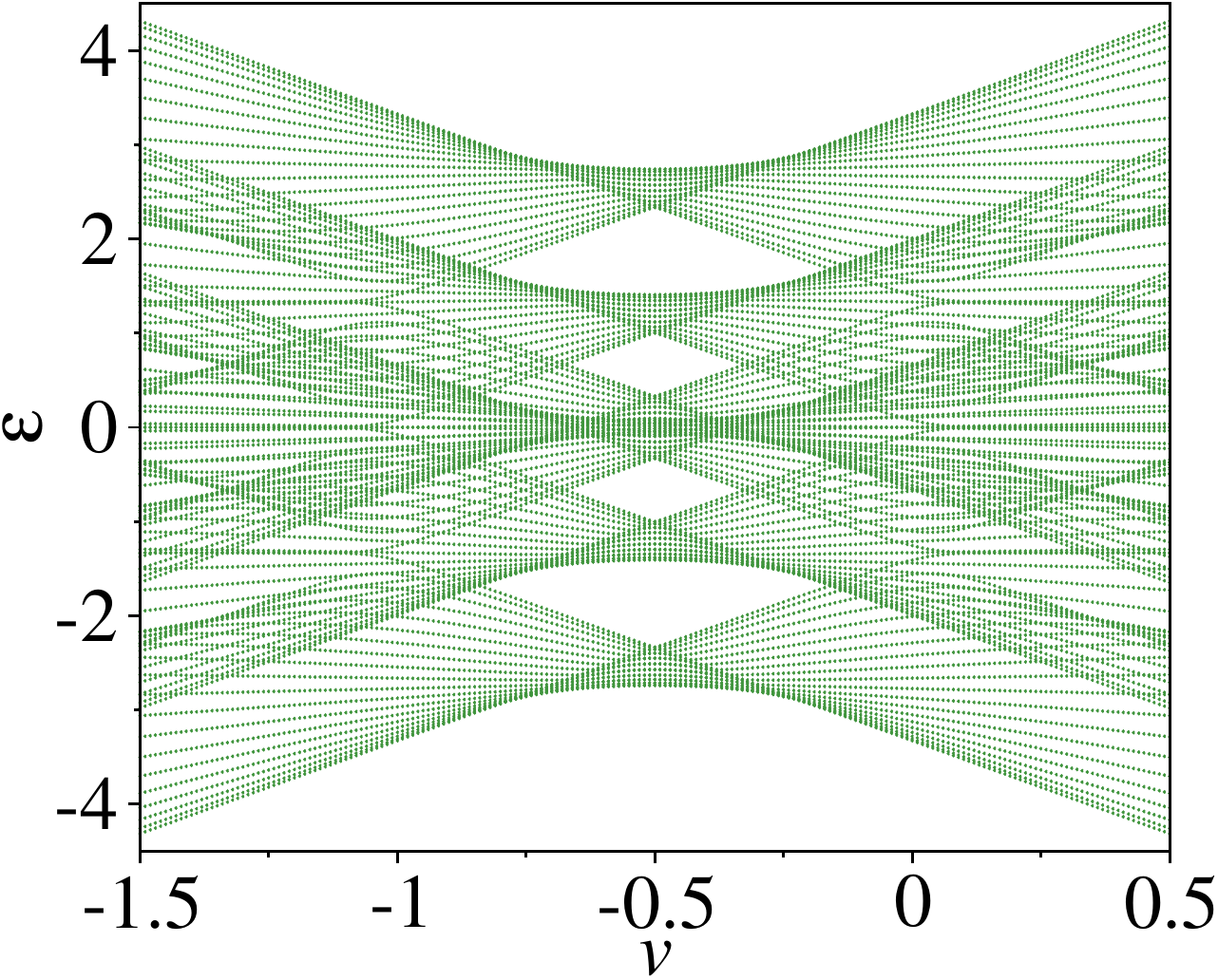}}\\
\subfloat[]{\includegraphics[height=6.5cm]{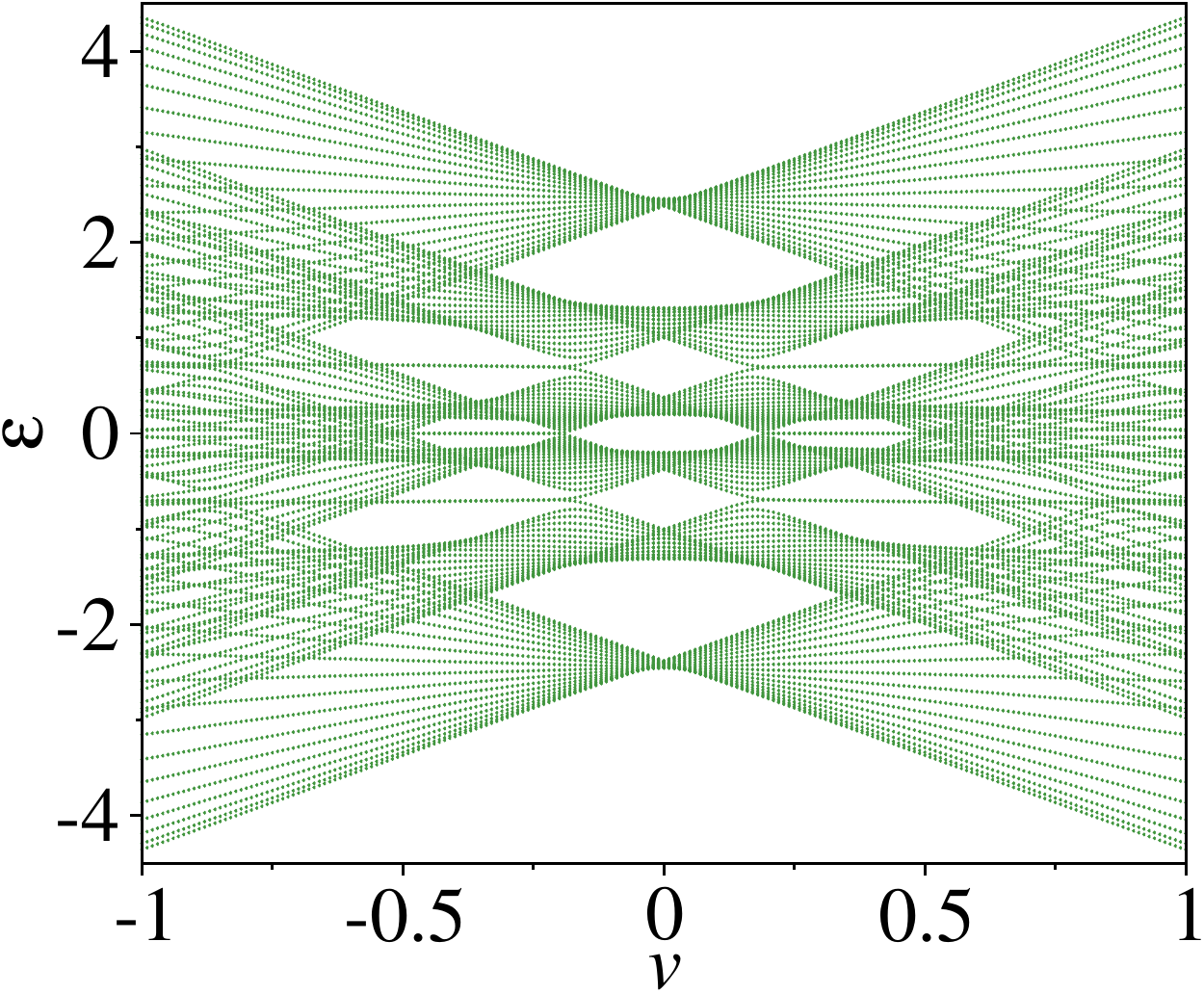}}
\subfloat[]{\includegraphics[height=6.5cm]{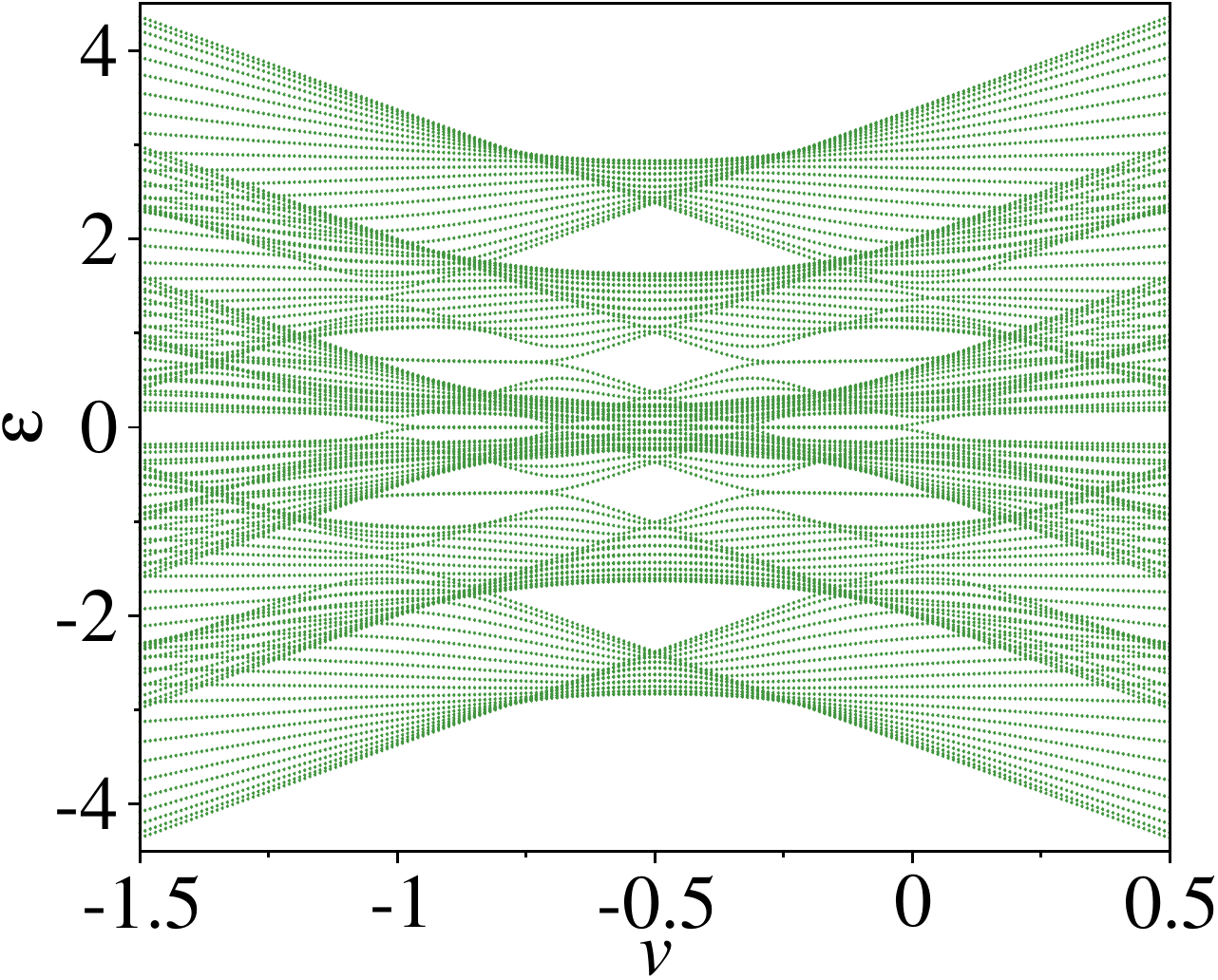}}
    \caption{Emergence of non equilibrium edge states in the quasi energy spectrum for driving frequency $\hbar\Omega=4c/3$ fixing $c=1$. In panels a-b (c-d) we show the undriven (driven) regime $g=0$ ($g=0.25c$), as a function of the intracell parameter $v$, with the intercell parameter satisfying $w/c=v/c$, and $w/c+1=v/c$ for (a),(c) and (b),(d) respectively.}
    \label{fig:my_label1}
\end{figure}
Within this context, the interpretation of the winding number is given in terms of the loop in momentum space associated to the reciprocal space representation of the bulk Hamiltonian of the system. As it is described in the following section, it is calculated to fully characterize any emergent topological phases. 

\noindent Once the driving protocol is turned on, the Hamiltonian evolve to a configuration showing zero quasi energy values, associated to edge states in a different $v$-range, as depicted in the lower panels (c) and (d) of \figurename{2}. In the next section we show that the configuration $v = w = -0.3, c = 1$ that presents edge-localized states, as shown in \figurename{2}-d, coincides with avoiding crossings in the bulk spectrum. We also observe that, as expected, not only zero energy  edge states appear in the quasi energy spectrum but additional ones emerge which are associated to the periodic nature of the quasi energy spectrum. Therefore, as we turn on the perturbation, edge-localized states developed in different energy levels because of the coupling between the Floquet replicas. Moreover, the driven regime exhibits the ubiquitous bandgaps induced by the time dependent perturbation. In order to get physical insight into the nature of the emergence of these high frequency modulation effects, we now analyze the bulk Hamiltonian and discuss the pseudospin polarization associated to the differences in the occupation of the bipartite system. 

\subsection{Bulk Hamiltonian}
Once we have numerically explored the quasi energy spectrum dynamics for the finite-size system and we have obtained the interesting result that zero energy edge states emerge in the driven scenario, we explore the dynamical consequences for the bulk properties, as given via an effective bulk Hamiltonian. For this purpose, we begin by remembering that, for one chain we have the effective $2\times2$ Hamiltonian in momentum space:
\begin{equation}
H_{1}=
\left(
\begin{array}{cc}
     0 & ve^{-ika} + w  e^{ika} \\
     ve^{ika} + w  e^{-ika} & 0
\end{array}    
\right)
\end{equation}
As it is schematically shown in the panel (a) of \figurename{1}, $v$ ($w$) describes the intra cell (inter cell) energy coupling within the single chain whereas $a$ is the lattice spacing, which from now on is set as energy scale (i.e. $a=1$), such that momenta are measured in units of $a^{-1}$. As shown in \figurename{1}, the layers are coupled via an effective term of strength $c$, which leads to the bulk Hamiltonian \cite{li2017topological}:
\begin{equation}\label{h0}
H_0=
\left(
\begin{array}{cc}
     0 &  v e^{-ik}+ w  e^{ik}  + c \\
     v e^{ik} +w  e^{-ik} +  c & 0
\end{array}    
\right)
\end{equation}
\begin{figure}\label{figure3}
    \subfloat[]{\includegraphics[height=6.5cm]{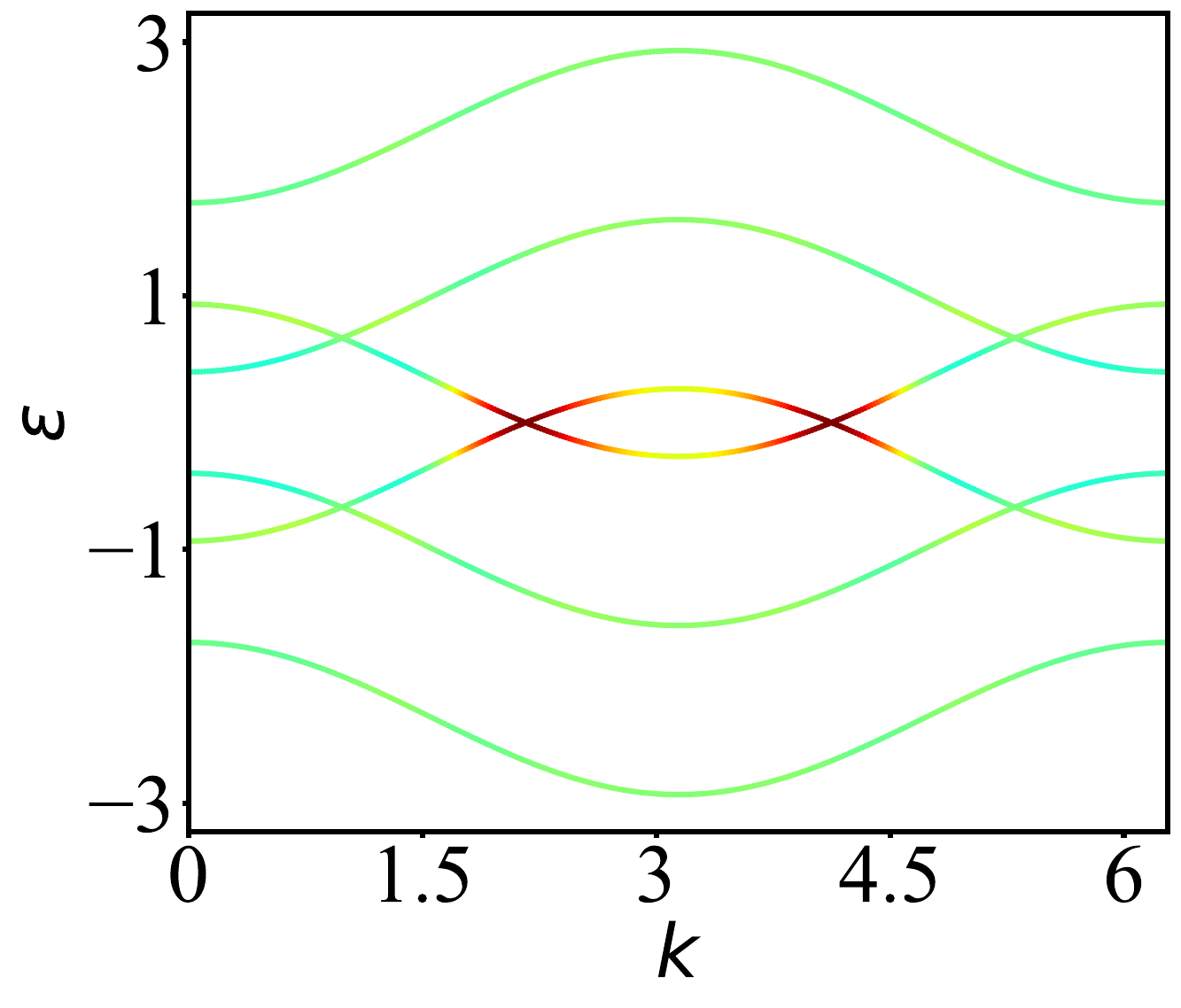}}
    \subfloat[]{\includegraphics[height=6.65cm]{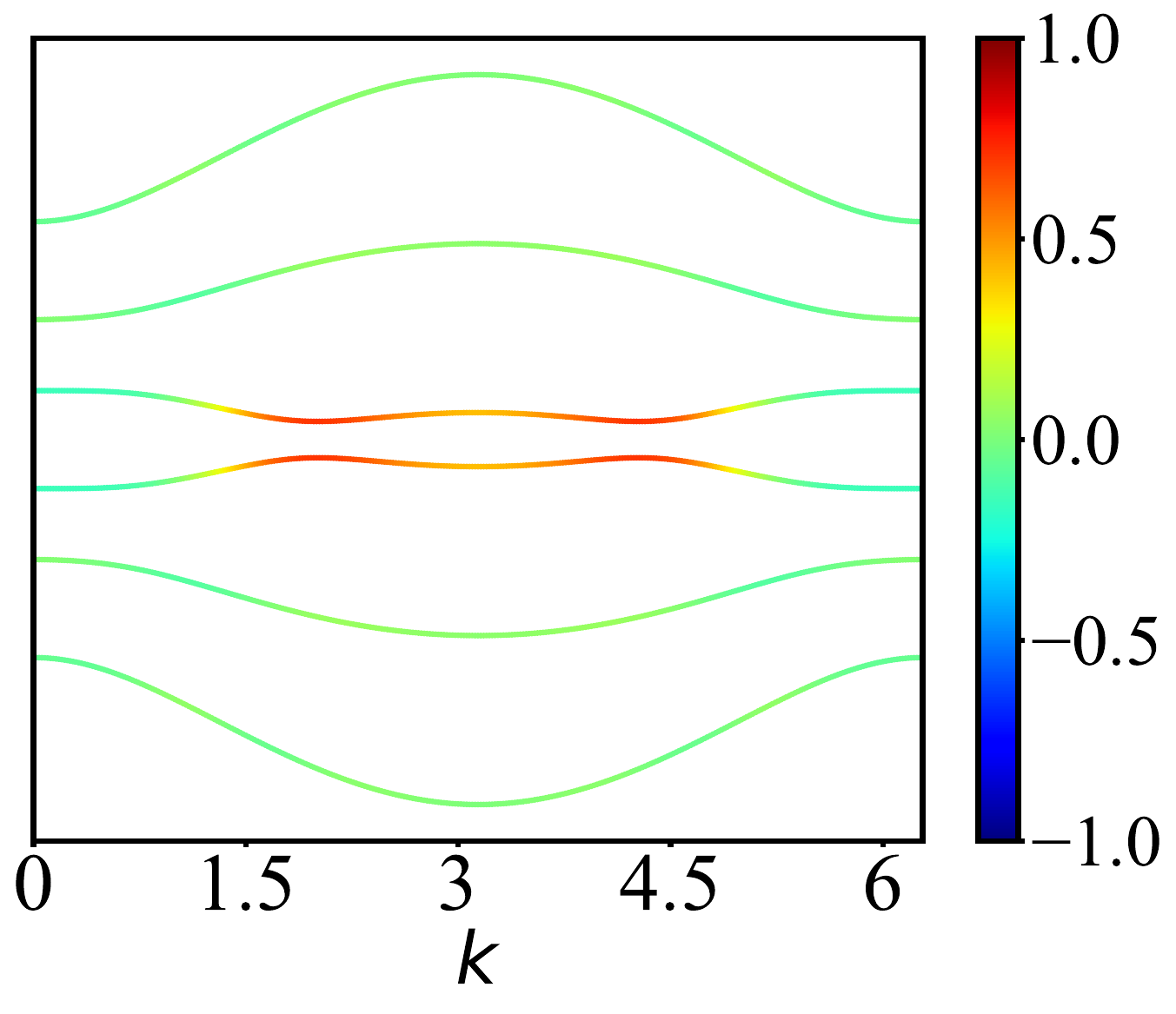}}
    \caption{Momentum dependence of the bulk quasi energy spectra for the undriven $g=0$ (panel a) and driven $g=0.25c$ (panel b) regimes. We have set the parameters as $v=w=0.3c$, $c=1$. The ubiquitous photoinduced bandgaps are present in panel b, and the color code represents the value of the mean pseudospin polarization.} 
\end{figure}
The bulk energy spectrum of the static Hamiltonian reads
\begin{equation}
\varepsilon_{0,s}=s\sqrt{[(v+w)\cos k+c]^2+[(v-w)\sin k]^2}    
\end{equation}
whereas its eigenstates can be written as
\begin{equation}
\ket{\Psi_s(k)}=\frac{1}{\sqrt{2}}
\begin{pmatrix}
1\\
s e^{i\beta_0}
\end{pmatrix},
\end{equation}
with 
\begin{equation}
\tan\beta_0=\frac{(v-w)\sin k}{(v+w)\cos k+c}.    
\end{equation}
From the eigenstates, the winding number can be calculated as
\begin{equation}
Z_0=\frac{i}{2\pi}\int^\pi_{-\pi}\bra{\Psi_s(k)}\partial_ k\ket{\Psi_s(k)}dk
\end{equation}
which is equivalent to
\begin{equation}
Z_0=-\frac{1}{2\pi}\int^\pi_{-\pi}\frac{\partial\beta_0}{\partial k}dk,
\end{equation}
which reproduces the results of the phase diagram given in reference \cite{li2017topological}.
The driving interaction in the bulk Hamiltonian takes the form  \cite{dal2015floquet}
\begin{equation}\label{interaction}
V(t) = 2g\cos(\Omega  t)
\left(
\begin{array}{cc}
     0 & -i\sin k  \\
     i\sin k & 0 
\end{array}
\right)
= 2g\cos(\Omega  t)\sigma_{y} \sin(k) 
\end{equation}
\noindent where $\sigma_{y}$ is the corresponding Pauli matrix in the canonical representation where $\sigma_{z}$ is diagonal. In this manner, we get the periodic time-dependent Hamiltonian 
\begin{equation}
    H(t) = H_{0} + V(t).
\end{equation}
The full numerical solution of the intermediate frequency regime leads to the results shown in figure \figurename{3} for either trivial and nontrivial topological configurations. As expected for periodically driven systems, the bulk quasi energy spectrum shows bandgaps which are due to the coupling among the different harmonics or sidebands. The color code represents the weight of the bulk pseudospin contribution with darker (lighter) zones corresponding to the maximum (minimum) value of the average pseudospin polarization. It is apparent that, for each configuration one can achieve maximum polarization at either the center or the boundary of the first Brillouin zone, since there is a largest weight of the polarization which can be used in order to select polarization configurations suitable for different momenta.\\ 
\noindent In order to get more physical insight, we make further analytical progress by constructing an effective Floquet Hamiltonian to leading order in the driving strength, as it is explicitly shown below at different frequency regimes.
\subsection{High frequency limit: Analytical Solution}\label{sectionhigh}
The high frequency regime allows for an exact analytical treatment of the driven problem since the Floquet sidebands get decoupled and the effects of the driving interaction can be effectively captured via an energy renormalization of the static energy spectrum. After some lengthy calculations (presented in the appendix), and setting $\xi=2g\sin k/\hbar\Omega$, we get the effective high frequency Floquet Hamiltonian:
\begin{equation}\label{highfrequency}
    H_{F} = (v-w)\sin k\sigma_{y} + [(w + v)\cos k + c] J_{0}(\xi)\sigma_{x},
\end{equation}
where $J_0(\xi)$ is the zeroth order Bessel function of the first kind.
The approximate quasi energy spectrum reads then 
\begin{equation}
\varepsilon_s=\sqrt{[(v+w)\cos k+c]^2 J^2_0(\xi)+(v-w)^2\sin^2 k}
\end{equation}
for which the Floquet states are given by
\begin{equation}
\ket{\phi_s(k)}=\frac{1}{\sqrt{2}}
\begin{pmatrix}
1\\
s e^{i\beta}
\end{pmatrix},
\end{equation}
with 
\begin{figure}\label{polarization1}
\includegraphics[width=8cm]{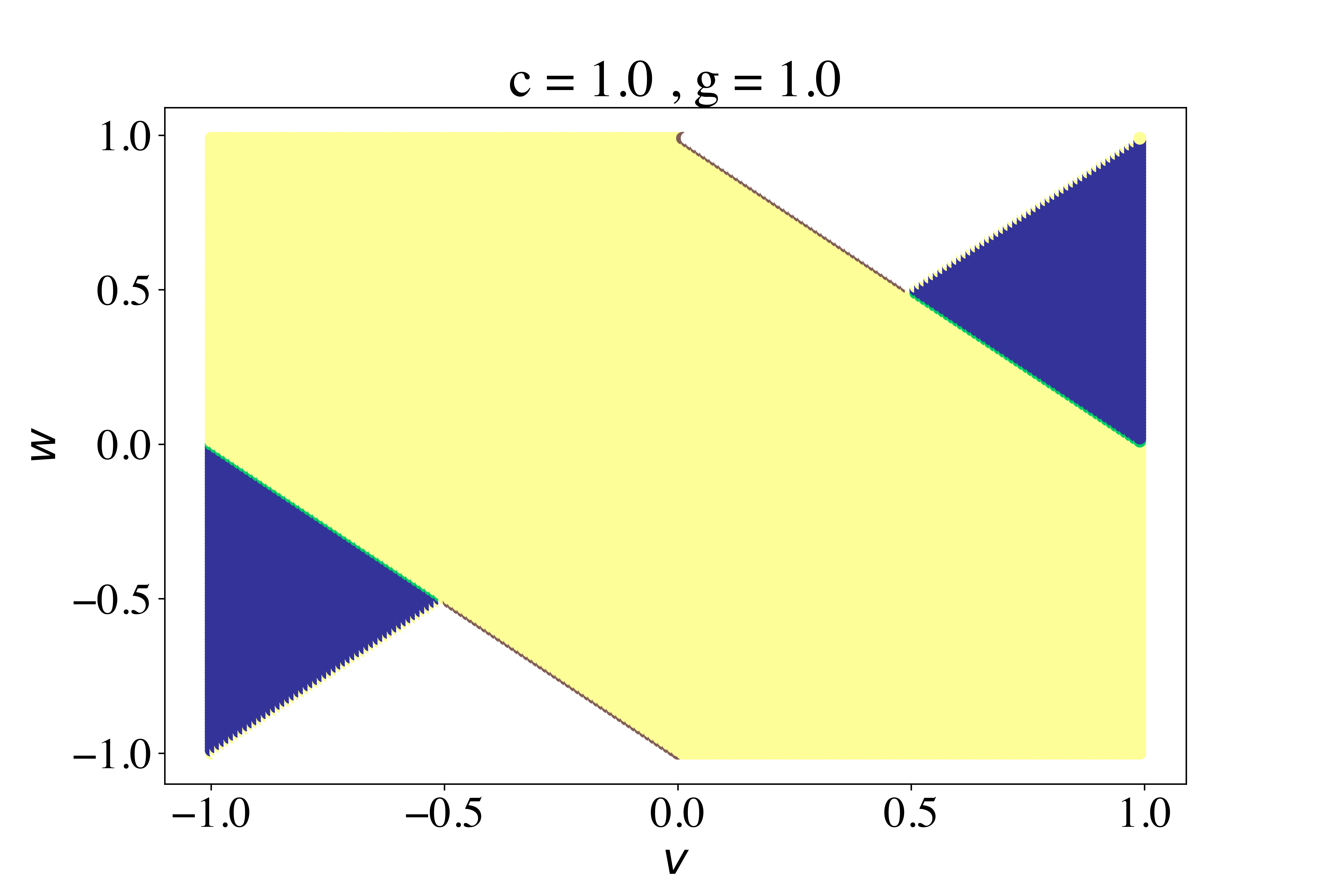}
\includegraphics[width=8cm]{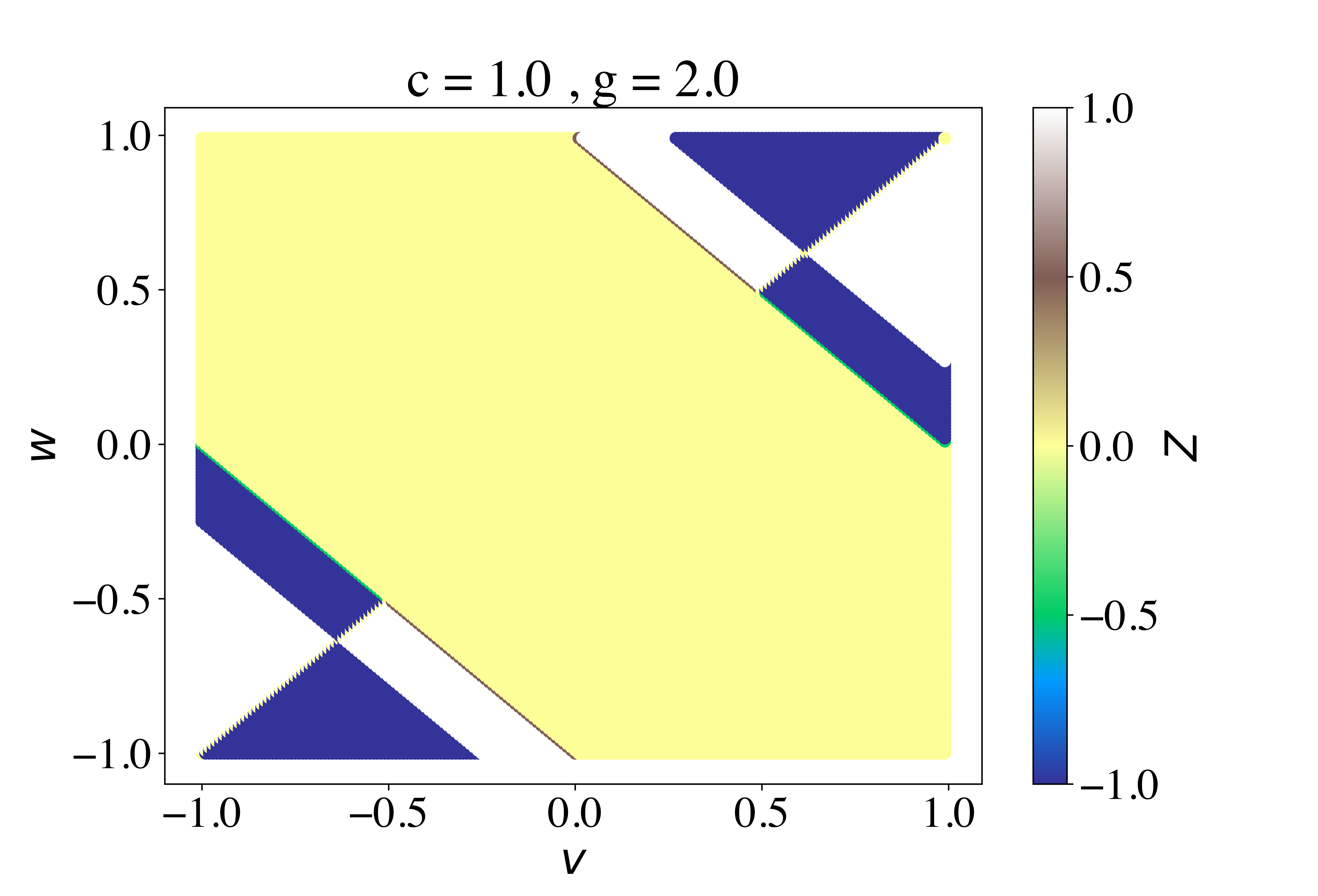}
\caption{Phase diagram for the winding number, within the effective high frequency Floquet Hamiltonian. The parameters are chosen as $c=1$, $g=c$ (left panel) and $g=2c$ (right panel). Alternating trivial and nontrivial phases are achieved for large values of the driving strength.}
\end{figure}
\begin{equation}
\tan\beta=\frac{\tan\beta_0}{J_0(\xi)}=\frac{(v-w)\sin k}{J_0(\xi)[(v+w)\cos k+c]}    
\end{equation}

\noindent We can then assess the topological effects of the perturbation by numerically evaluating the winding number
\begin{equation}
Z=\frac{1}{2\pi}\int^\pi_{-\pi}dk\frac{d\beta}{dk},    
\end{equation}
which is shown in \figurename{4}, in the regime where the two chains are strongly coupled $c=w=1$, for effective parameter $g=c$ (left panel) and $g=2c$ (right panel). First of all, we observe in the right panel that a phase inversion is induced at large values of the effective coupling, which exemplifies the feasibility of the driving protocol to induce phase transitions which has been shown to be attained by monochromatic laser irradiation. 
Strikingly, upon analyzing the behaviour of the winding number along the line $c=w+v$ it is interestingly found that additional $Z=\pm\frac{1}{2}$ arise (see right panel on \figurename{4}). This apparently paradoxical result emerges by explicit evaluation of the parameters prior to integration and it is already present in the static domain ($\xi\rightarrow0$)
\begin{equation}
Z_0= \frac{1}{2\pi}\int^\pi_{-\pi}\frac{d}{dk}\left[\arctan{\left(\frac{(v-w)\sin k}{c+(w+v)\cos k}\right)}\right]dk    
\end{equation}
setting $c=v+w$, for $w=0$ one gets
\begin{equation}
\frac{1}{2\pi}\int^\pi_{-\pi}\frac{d}{dk}\left[\arctan{\tan k/2}\right]dk    
\end{equation}
which equals $1/2$.
\begin{figure}
    \centering
    \includegraphics[width=8cm]{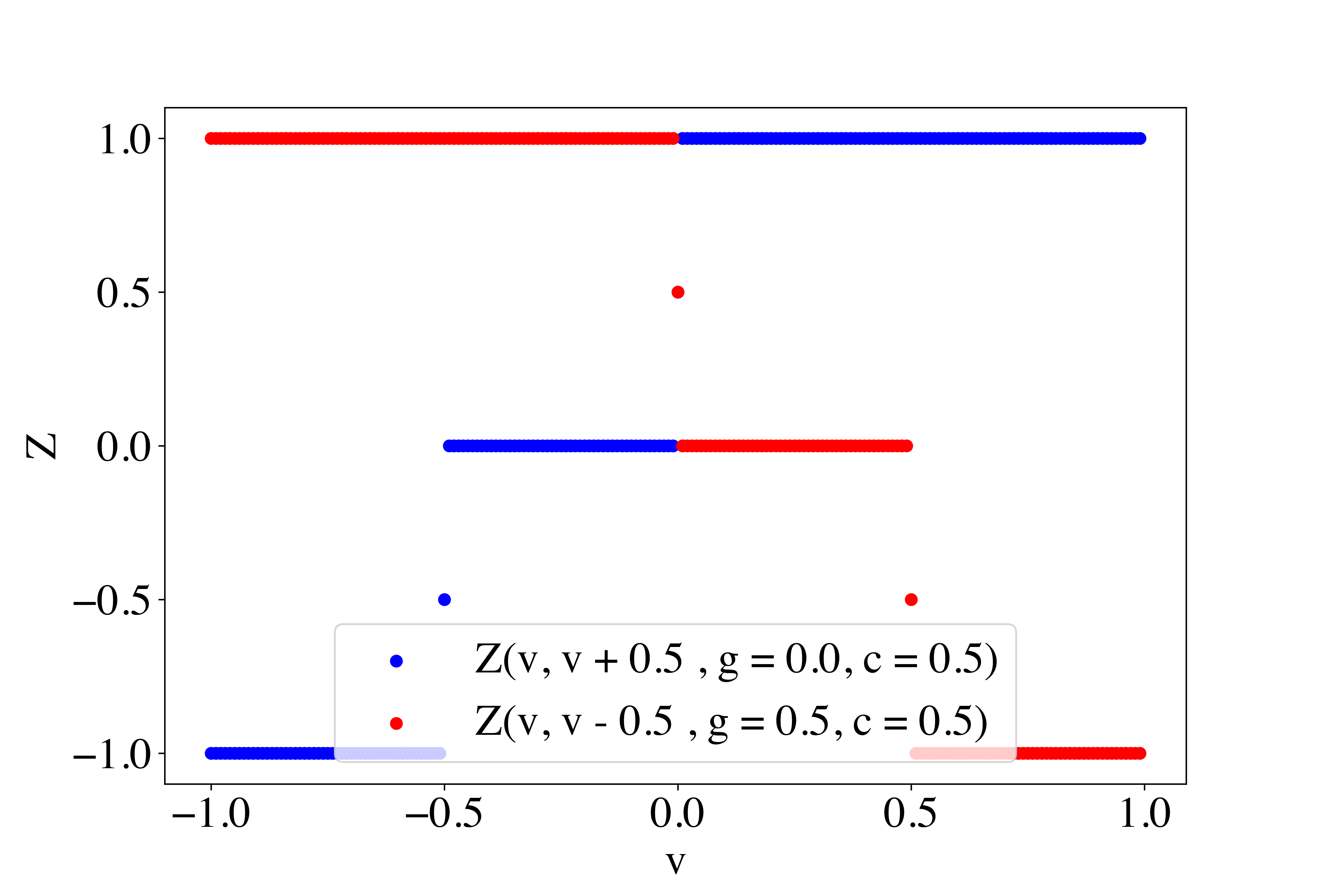}
    \includegraphics[width=8cm]{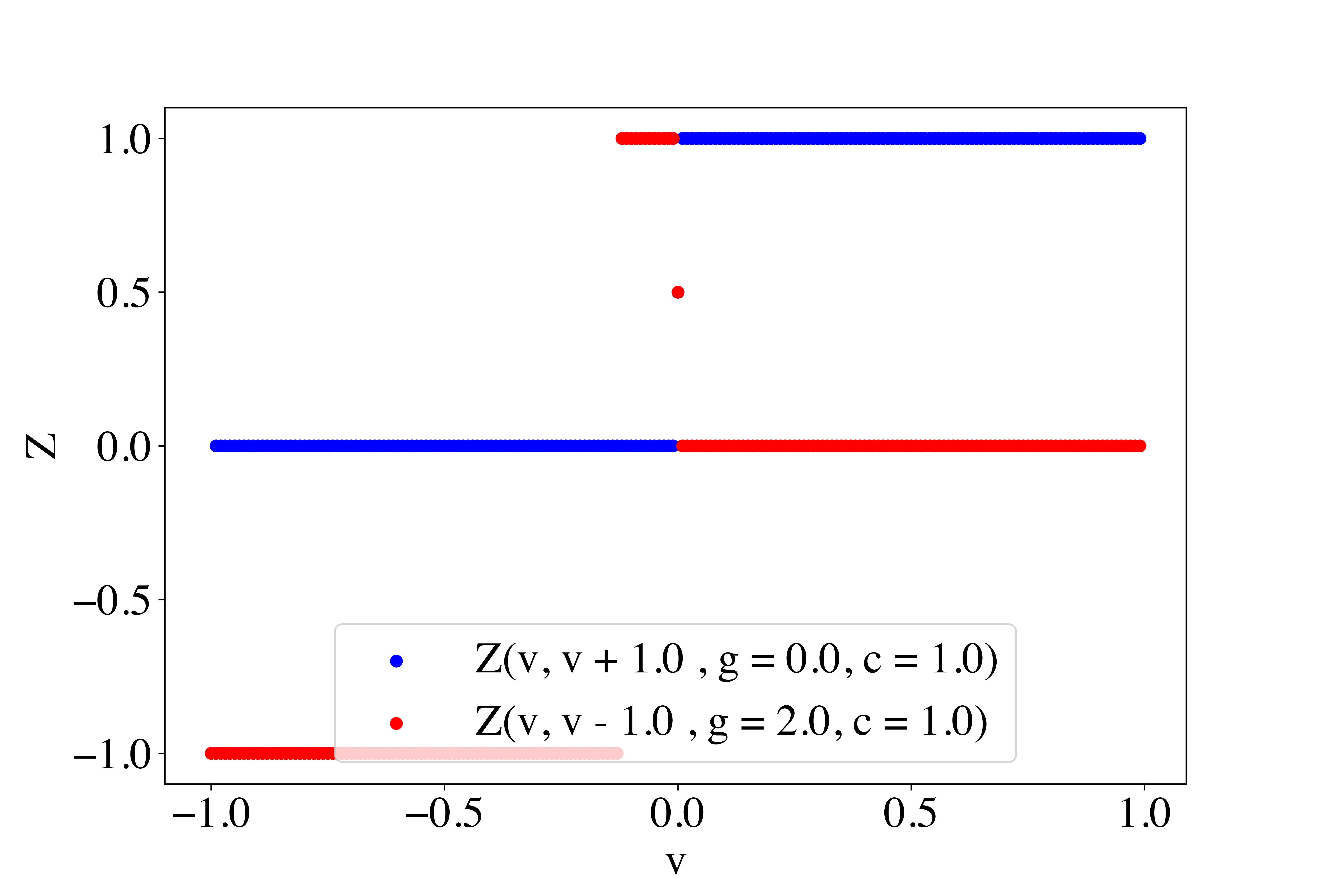}
    \caption{Winding number for the undriven and driven scenarios. At low interchain coupling values (left panel) $Z=\pm1/2$ non integer value emerges, whereas the strongly coupled chains configuration shows and isolated $Z=+1/2$ phase.}
    \label{fig:my_label}
\end{figure}
Although some theoretical works have put forward the idea of non integer winding number in non hermitian systems and Ising systems with long range correlations, and taking into account that these peculiar values happen at the transition between two integer and well defined phases, their understanding might require an experimental realization that could establish the physical implications of the actual emergence of this apparently paradoxical result.  

\noindent Having focused on the winding number and topological phases, we can also  evaluate another physical quantities of interest as the pseudospin polarization, measuring the population imbalance among the two the bipartite sublattices. For this purpose, we consider the system to be prepared in an initial state $\ket{\psi_{k}}$ at $t=0$. Thus,  the associated average lattice polarization can be deduced from the definition
\begin{eqnarray}
    \braket{\sigma_{z}} &=& \frac{1}{T} \int_{0}^{T} \bra{\psi_k}U^{\dagger}(t)\hat{\sigma_{z}}U(t)\ket{\psi_k} dt \notag.
\end{eqnarray}
Choosing the system to be initially prepared in an eigenstate of the static Hamiltonian, given in equation (\ref{h0}), one finds the result (the details of the calculation are deferred for the appendix) 
\begin{equation}\label{polar}
        \langle \sigma_{z}\rangle  =-s\sin(\beta-\beta_{0})\sin(2\pi\Sigma_{k})\sinc(2\pi \Sigma_{k}),
\end{equation}
where $s=\pm1$ and $\Sigma_k=\varepsilon_k/\hbar\Omega$.    
One then finds an oscillating pattern in the average polarization which is modulated via the phase difference $\beta - \beta_{0}$ of the dressed and static eigenstates, respectively. In this manner, the role of the driving interaction is to generate a non-vanishing average value for the pseudospin polarization. This can explicitly be seen in \figurename{6} where we plot equation (\ref{polar}), for both $w/v<1$ and $w/v>1$ configurations. Remarkably, from equation \ref{polar} in either case, the polarization vanishes in the absence of the driving protocol. Moreover, this non-vanishing of the $z$ component of the pseudospin degree of freedom leading to driven pseudospin oscillations has the physical importance of being connected to the chiral symmetry breaking  which is absent in the static regime. Thus, although from this polarization we can not directly infer the topological phase of the system we can signal the violation of the chiral symmetry needed for the system to remain in such topological state.
\begin{figure}\label{polarization2}
\includegraphics[width=8cm]{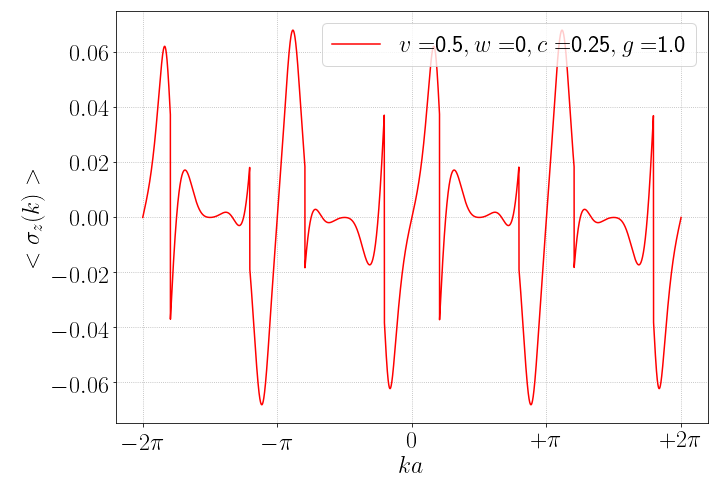}     
     \includegraphics[width=8cm]{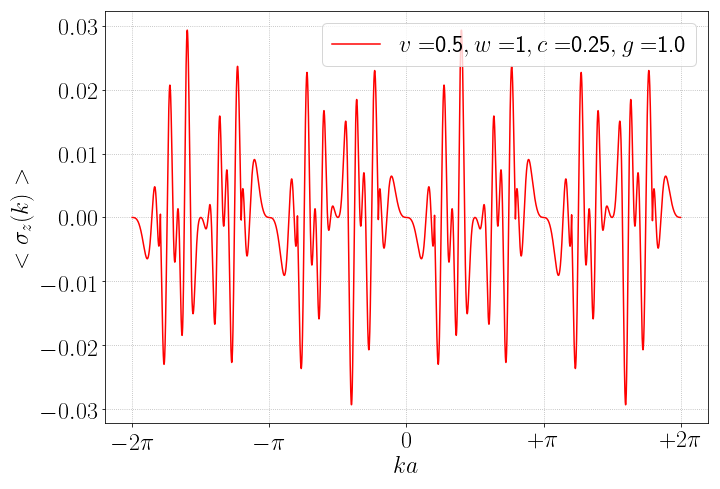}
\caption{Average value of the pseudospin polarization as given in equation (\ref{polar}), where we have chosen the $s=-1$ eigenstate of the static Hamiltonian. The left (right) figure shows the polarization for the  $w/v<1$ ($w/v>1$) configurations.}
\end{figure}
\section{Discussion and conclusions}\label{section3}
In this work we have analyzed the nonequilibrium topological phases in the two coupled SSH chain system to discern how the interplay of interchain coupling and driving interaction in the high frequency regime may lead to nontrivial dynamical behaviour. We have explored the emergent properties of two coupled one-dimensional SSH chains subject to a periodically driving perturbation that drives the system out of equilibrium. The exact numerical results in a finite size chain show that emergent zero energy states appear in the quasi energy spectrum for parameter values that, within the static regime, are absent. The reason for the appearance of these additional edge states is the contribution arising from the sidebands which modify the topology of the static energy bands for the static Hamiltonian. In order to give an effective description of the underlying physical processes that take place in the nonequilibrium regime, we have studied the bulk properties of the system. As expected, we have found that leading order bandgap contributions arise from the lower energy sidebands which, in the high frequency regime, can renormalize the static energy band spectrum or lead to band mixing that induces bandgaps responsible for the topologically non-trivial response of the system in the intermediate frequency regime. 
As it was reported in reference \cite{Platero2013}, we also find that alternating phases can be achieved by tuning the effective strength of the driving protocol. We would expect that these results could be experimentally probed in the realm of ultracold atoms in optical lattices by a combination of Bloch oscillations and Ramsey interferometry\cite{Nature2013_Kitawaga} or by using phase-sensitive injection, and adiabatic preparation schemes as reported in reference \cite{Meier2016}. Strikingly, we find a configuration leading to a non integer value for the winding number, which happens at one of the trivial-nontrivial transition. Interestingly, the possibility for non integer values of the winding number has been described recently in reference\cite{noninteger}. In their work, the authors propose a generalization of the definition of this topological invariant to make sense of the anomalous situations for which the winding number is not well defined. Moreover, a recent experimental work\cite{nonintegerexp}, explores the physical feasibility of realizing these potentially interesting non integer winding numbers in the skyrmion texture present in \text{$Cu_2OSeO_3$}. The authors determine that although these noninteger winding number values might be experimentally detected, they appear to be energetically unstable and thus difficult to measure via their circular dichroism technique. In addition, these non integer winding numbers have also been recently predicted in static non hermitian materials\cite{nonhermitian}. In their work, the authors introduced a complex Berry phase, which is used as a quantized invariant to predict the existence of gapless edge modes in these static non hermitian systems. The extension of driving protocols to these non dissipative Hamiltonians, suggesting the generation of new types of Floquet topological materials has been recently put forward\cite{drivennh}. The authors discuss and compare the skin effects in non Bloch band theory and their emergent non Floquet system, which might find application in designing dissipative photonic topological materials. Therefore, we might expect further interesting developments dealing with the interplay of coupled chains, described by non hermitian models, that could be explored elsewhere.   
\section{appendix}\label{appendix}
Here we present the details of the calculations for the relevant results discussed in the main text. 
\subsection{Derivation of the high frequency effective Floquet Hamiltonian} 
Starting from the periodic time-dependent Hamiltonian
\begin{equation}
H(t) = H_0 +V(t)
\end{equation}
with $H_0$ and $V(t)$ given by equations \ref{h0} and \ref{interaction}, respectively we explicitly show how to derive the effective high frequency Hamiltonian given in equation \ref{highfrequency}. In order to do so, we first perform a unitary transformation
\begin{equation}
    \tilde{H}(t) = P^{\dagger}(t)H(t)P(t) - i\hbar P^{\dagger}(t) \partial_{t}P(t)
\end{equation}
with the definition $P(t) = e^{\frac{i}{\hbar}\int V(t) dt}$, which in turn leads to
\begin{equation}
    \tilde{H}(t) = e^{iM(t)\sigma_{y}} H(t)e^{-iM(t)\sigma_{y}} -V(t)
\end{equation}
where we have defined
\begin{equation}
    M(t)=\frac{1}{\hbar} \int f(t) dt \sin k = 2g\sin k \frac{\sin(\Omega t)}{\hbar\Omega }.
\end{equation}
Therefore, we get 
\begin{eqnarray}
\tilde{H}(t) &=& [\tilde{H}(t) - V(t)] + e^{iM(t)\sigma_{y}} [H(t), e^{-iM(t)\sigma_{y}}] \notag\\
&=& H_{0} + e^{iM(t)\sigma_{y}} [H_{0}, -i\sigma_{y}]\sin M(t) \notag\\
&=& H_{0} + (\cos M(t) + i\sin M(t)\sigma_{y})[((w  + v)\cos k + c)\sigma_{x}, -i\sigma_{y}]\sin M(t) \notag\\
&=& H_{0} + [(w  + v) \cos k + c]e^{iM(t)\sigma_{y}}\sigma_{z} 2 \sin M(t) 
\end{eqnarray}
\noindent The last term can be worked out as follow:
\begin{eqnarray}
    2 e^{iH(t)\sigma_{y}}\sigma_{z}\sin M(t) &=& 2(\cos M(t)\mathbb{1} + i\sin M(t) \sigma_{y})\sigma_{z} \sin M(t) \notag\\
    &=& \sin 2M(t)\sigma_{z} - (1+\cos 2M(t))\sigma_{x}\notag\\
    &=& -\sigma_{x} + \sin 2M(t) \sigma_{z} + \cos 2M(t)\sigma_{x}
\end{eqnarray}
which $\rightarrow 0$ as $M(t)$ (i.e. $g$) $\rightarrow 0$. Therefore,
\begin{equation}
    \tilde{H}(t) = H_{0} + [(\Omega  + v)\cos(ka) +c]\sigma_{x} + [\sin 2M(t)\sigma_{z} + \cos 2M(t) \sigma_{x}][(v+w)\cos k + c]
\end{equation}
At high frequencies we find to leading order $\sin 2M(t)\approx0$ and $\cos 2M(t)\approx J_0(\xi)$, with $\xi=\frac{2g\sin ka}{\hbar\Omega}$ and $J_0(x)$ being the zeroth order Bessel function. Thus, we get the effective Floquet Hamiltonian
\begin{equation}
    H_{F} = (v-w )\sin(ka)\sigma_{y} - [(\Omega  + v)\cos(ka) + t] J_{0}(\xi)\sigma_{x}.
\end{equation}
\subsection{Derivation of the pseudspin polarization}
Here, we start from the definition of the average pseudospin polarization
\begin{eqnarray}
    \braket{\sigma_{z}} &=& \frac{1}{T} \int_{0}^{T} \bra{\psi_{k}}U^{\dagger}(t)\hat{\sigma_{z}}U(t)\ket{\psi_{k}} dt \notag\\
    &=& \frac{1}{2\pi}\int_{0}^{2\pi}\bra{\psi_{k}}U^{\dagger}(\theta)\hat{\sigma_{z}}U(\theta)\ket{\psi_{k}}d\theta; \qquad \theta = \Omega  t
\end{eqnarray}
where for simplicity we can choose
\begin{equation}
\ket{\psi}=\left(
\begin{array}{cc}
     0 \\
     1 
\end{array}
\right)
\end{equation}
Notice that, within our effective Floquet Hamiltonian,
\begin{equation}
    U(t)= e^{-i H_{k}t / \hbar}= e^{-i\Sigma_{k}\sigma_{k}\theta} = U(\theta)
\end{equation}
where 
\begin{equation}
\begin{split}
\Sigma_{k}\sigma_{k}  
& =\frac{1}{\hbar \Omega } \left[(v-w ) \sin(ka) \sigma_{y}-J_{0}(\xi)\left[(v+w )\cos(ka)+c\right ]\sigma_{x}  \right] \\
& = \Sigma_{k} \begin{pmatrix}
0 & e^{-i\beta} \\ 
e^{i\beta} & 0
\end{pmatrix}
\end{split}
\end{equation}
with
\begin{equation}
   \tan\beta=-\frac{(v-w )\sin(ka)}{J_{0}(\xi)\left [(v+w )\cos(ka)+c\right] }
\end{equation} 
and  
\begin{equation}
    \Sigma_{k}=\frac{1}{\hbar \Omega }\sqrt{(v+w )^{2}\sin^{2}(ka)+ J_{0}^{2}(\xi)\left[(v+w )\cos(ka)+c\right]}= \frac{\varepsilon_{k}}{\hbar\Omega }
\end{equation}
Therefore
\begin{equation}
    U(\theta)= \cos(\Sigma_{k}\theta)\mathbb{1}-i\sin(\Sigma_{k}\theta)\sigma_{k}
\end{equation}
which leads to
\begin{equation}
\begin{split}
   U^{\dagger}(\theta)\sigma_{z}U(\theta) & = \begin{pmatrix}
\cos\theta &i\sin\theta e^{-i\beta} \\ 
 i\sin\theta e^{i\beta}& \cos\theta
\end{pmatrix}
\begin{pmatrix}
1& 0\\ 
0& -1
\end{pmatrix}
\begin{pmatrix}
\cos\theta & -i\sin\theta e^{-i\beta} \\ 
 -i\sin\theta e^{i\beta}& \cos\theta
\end{pmatrix} \\
&= \begin{pmatrix}
\cos\theta &-i\sin\theta e^{-i\beta} \\ 
 i\sin\theta e^{i\beta}& -\cos\theta
\end{pmatrix}
\begin{pmatrix}
\cos\theta & -i\sin\theta e^{-i\beta} \\ 
 -i\sin\theta e^{i\beta}& \cos\theta
\end{pmatrix} \\
& = \begin{pmatrix}
\cos2\theta &i\sin2\theta e^{-i\beta} \\ 
 i\sin2\theta e^{i\beta}& -\cos2\theta
\end{pmatrix} \\
& = \begin{pmatrix}
\cos(2\Sigma_{k}\theta) & -i\sin(2\Sigma_{k}\theta)e^{-i\beta} \\ 
i\sin(2\Sigma_{k}\theta)e^{-i\beta}& -\cos(2\Sigma_{k}\theta)
\end{pmatrix}
\end{split}
\end{equation}
Upon substitution on the relation $\left \langle \psi_{k}|U^{\dagger}(\theta)\sigma_{z} U(\theta)|\psi_{k} \right \rangle$ we get
\begin{equation}
\begin{pmatrix}
0 & 1 
\end{pmatrix}
\begin{pmatrix}
-i\sin(2\Sigma_{k}\theta) e^{-i\beta} \\
-\cos(2\Sigma_{k}\theta)
\end{pmatrix} = -\cos(2\Sigma_{k}\theta)
\end{equation}
Thus, we only had to evaluate
\begin{equation}
    \begin{split}
        \left \langle \sigma_{z}(k) \right \rangle & = \frac{-1}{2\pi} \int_{0}^{2\pi} \cos\left (2\frac{\varepsilon_{k}}{\hbar \omega }\theta\right) d\theta = \frac{\sin \left (2\frac{\varepsilon_{k}}{\hbar \Omega }\theta\right)}{2\pi \left (2\frac{\varepsilon_{k}}{\hbar \Omega }\theta\right)}\Big|_{0}^{2\pi} \\
        & = \sinc \left [ \frac{4\varepsilon_{k}\pi}{\hbar\Omega }\right]
    \end{split}
    \label{spin-eq}
\end{equation}
Another interesting scenario arises when the system is initially prepared in an eigenstate of the static Hamiltonian
\begin{equation}
    H_{0}=\varepsilon_{0k} \begin{pmatrix}
    0 & e^{-i\beta_{0}} \\
    e^{i\beta_{0}} & 0 
    \end{pmatrix}
\end{equation}
where 
\begin{equation}
    \begin{split}
        \varepsilon_{0k} & = \sqrt{[(v+w)\cos(ka) +c]^2 + (v-w )^{2}\sin^{2}(ka) }\\
        %& = \sqrt{v^{2}+\omega ^{2}+2v\omega  \cos(2ka)+ c^{2}\cos^{2}(ka)+2(v+\omega )c \cos^{2}(ka)}
    \end{split}
\end{equation}
and
\begin{equation}
tan\beta_{0}=\frac{(v-w ) \sin(ka)}{(v+w )\cos(ka)+c}
\end{equation}
The corresponding eigenstates read
\begin{equation}
    |\psi_{\nu}  \rangle = \frac{1}{\sqrt{2}} \begin{pmatrix}
    1 \\ \nu e^{i\beta_{0}}
    \end{pmatrix}
\end{equation}
with $\nu=\pm1$ such that $H_{0}|\psi_\nu\rangle =\nu\epsilon_{0k}|\psi_\nu\rangle$.
Thus, within this approximate scenario we get for the $\nu=-1$ state
\begin{equation}
    \begin{split}
        \langle \psi_{-1}| U^{\dagger}(\theta)\sigma_{z}U(\theta)|\psi_{-1}\rangle & =\frac{1}{2}\begin{pmatrix}
        1 & -e^{-i\beta_{0}}
        \end{pmatrix}\begin{pmatrix}
\cos(2\Sigma_{k}\theta) & -i\sin(2\Sigma_{k}\theta)e^{-i\beta} \\ 
i\sin(2\Sigma_{k}\theta)e^{-i\beta}& -\cos(2\Sigma_{k}\theta)
\end{pmatrix}\begin{pmatrix}
1 \\ -e^{i\beta_{0}}
\end{pmatrix} \\
& = \frac{1}{2} \begin{pmatrix}
        1 & -e^{-i\beta_{0}}
        \end{pmatrix}\begin{pmatrix}
\cos(2\Sigma_{k}\theta)+ i\sin (2\Sigma_{k}\theta)e^{-i(\beta-\beta_{0})}  \\ 
i \sin(2\Sigma_{k}\theta)e^{i\beta} + \cos(2\Sigma_{k}\theta)e^{i\beta_{0}}
\end{pmatrix}\\
& =-\sin(2\Sigma_{k}\theta)\sin(\beta-\beta_{0})
    \end{split}
\end{equation}
Thus, the result for the average  polarization reads now
\begin{equation}
   \begin{split}
        \langle \sigma_{z}\rangle  & = -\frac{1}{2\pi}\int_{0}^{2\pi}\sin(2\Sigma_{k}\theta)\sin(\beta - \beta_{0}) d\theta  \\
        & = \sin(\beta- \beta_{0}) \left[\frac{\cos(2\Sigma_{k}2\pi)-1}{2\pi 2\Sigma_{k}}\right]\\
        & =- \sin(\beta-\beta_{0})\sin(2\pi\Sigma_{k})\sinc(2\pi \Sigma_{k}) 
   \end{split}
\end{equation}
\acknowledgements{The authors thank R. Molina for useful discussions. 
%This work is part of the project: "Estudio de la fractalidad en sistemas electr\'onicos" at ESPOL.
}

\end{document}